\documentclass[twocolumn,english,aps,groupedaddress]{revtex4}
\usepackage[T1]{fontenc}
\usepackage[latin1]{inputenc}
\usepackage{graphicx}
\usepackage{amssymb}

\makeatletter

\providecommand{\tabularnewline}{\\}

\usepackage{graphicx}

\usepackage{babel}
\makeatother
\begin{document}

\title{g-factor engineering and control in self-assembled quantum dots}

\author{G. Medeiros-Ribeiro}

\email{gmedeiros@lnls.br}

\author{E. Ribeiro}

\author{H. Westfahl Jr.}

\affiliation{Laborat\'{o}rio Nacional de Luz S\'{\i}ncrotron, PO Box 6192, 13084-971
Campinas - SP, Brazil}

\date{\today{}}

\begin{abstract}
The knowledge of electron and hole g-factors, their control and engineering
are key for the usage of the spin degree of freedom for information
processing in solid state systems. The electronic g-factor will be
materials dependent, the effect being larger for materials with large
spin-orbit coupling. Since electrons can be individually trapped into
quantum dots in a controllable manner, they may represent a good platform
for the implementation of quantum information processing devices.
Here we use self-assembled quantum dots of InAs embedded in GaAs for
the g-factor control and engineering. 
\end{abstract}


\maketitle

\section{Introduction}

\label{intro}

The prescription for a quantum computer implementation proposed by
DiVincenzo \cite{1} consists of a series of stringent requirements
that have necessarily to be fulfilled. In particular, the optimum
conditions for a functional qu-bit, the basic unit for the representation
of the information in its quantum form, can only be found in very
few systems, most specifically those which allows one to address,
operate and assess the selected quantum degree of freedom. The capability
of storing electrons, one by one in a predictable and controllable
manner \cite{2}, makes self-assembled quantum dots (QDs) a promising
candidate for the use of the spin degree of freedom of trapped electrons
as a qu-bit. It has been shown that one can charge arrays of $1\times10^{8}$
QDs with one electron in a reproducible way, presenting also robustness
against thermal cycling \cite{3}. In order to produce such systems,
one needs to have an extremely uniform array of QDs, where the inhomogeneous
line broadening due to the size distribution is a smaller effect when
compared to the Coulomb charging energy.

Knowledge of the g-factor is paramount for any application that requires
the control of the spin degree of freedom of electrons and holes.
Therefore, to successfully implement spintronics into semiconductor
heterostructures, one has to achieve g-factor assessment, control,
and engineering. Traditionally, g-factors are measured via electron-spin
resonance (ESR) techniques \cite{4}, from which the g-factor as well
as the dephasing and relaxation times can be precisely evaluated.
The difficulty in applying ESR technique to ensembles of QDs lies
on the smallest number of spins that can be measured, which hovers
around $10^{12}$ spins for the lowest detection limit. Nonetheless,
using a variety of optical techniques, g-factors in quantum dots \cite{5,6,7}
have been evaluated. In lithographically defined QDs \cite{8} as
well as in self-assembled quantum dots \cite{3}, and in metallic
nanoparticles \cite{9} one can assess the g-factor by using transport
spectroscopy experiments. Transport spectroscopy is a powerful technique
that allows the assessment of the spin properties of a very small
number of electrons \cite{10}.

The control of the electronic g-factor has been demonstrated recently
in parabolic quantum wells, whereby applying an electric field one
can move the electron and hole wavefunctions in and out of different
g-factor materials \cite{11,12}. By tailoring the composition profile
of a given crystal, it is possible to engineer structures with different
g-factors in a similar way that band-gap engineering has been implemented
\cite{13}. This has important consequences for optoelectronic devices
that make use of the spin degree of freedom in hypothetical quantum
networks \cite{14}.

The purpose of this work is to assess the magnitude of the g-factor
of electrons trapped in the ground state of InAs self-assembled quantum
dots using capacitance spectroscopy, the external control of the g-factor,
and the demonstration of g-factor engineering in QDs.

\section{Sample growth and processing}

\label{sec:1}

The samples were grown by Molecular Beam Epitaxy, and consisted of
an undoped 1 $\mu$m thick GaAs buffer layer grown at 600 $^{o}$C,
followed by a back contact, Si doped nominally to $1\times10^{18}$
cm$^{-3}$ and 80 nm thick. The temperature was then lowered to 530
$^{o}$C, and an undoped 25 nm thick GaAs tunneling barrier layer
thickness (t$_{b}$) was grown, above which the InAs QD layer was
nucleated. During the InAs deposition the substrate was not rotated
resulting in a varying coverage throughout the wafer, therefore producing
a wealth of slightly different conditions in a single run. For sample
A, the amount of InAs deposited was about 1.9 ML, sufficient to form
$1-2\times10^{10}$ cm$^{-2}$ QDs. The structure was capped with
a 150 nm thick undoped GaAs layer (t$_{i}$). For the other samples
(B and C), a so-called strain reducing layer (SRL) \cite{15} In$_{0.2}$Ga$_{0.8}$As
30 \AA~ thick was deposited over the QD layer. In a similar fashion
the wafer had a spatially varying composition and thickness for the
SRL. Finally, a control sample (sample D) was grown exactly like sample
A, with the only difference being the way the InAs layer was deposited.
In this case, alternate beam epitaxy was utilized during the InAs
QD layer, alternating the As and In fluxes during the deposition process
at 0.1 ML intervals. The photoluminescence spectra of all samples
were measured and exhibited narrow lines ($\sim$40 meV for the ground
state transition) and shell structure up to 4 discrete levels of the
trapped carriers.

Figure 1a shows a 1 $\mu$m $\times$ 1 $\mu$m Atomic Force Microscopy
(AFM) image of an uncapped sample, and figure 1b a Cross Sectional
Transmission Electron Microscopy (XTEM) image of a buried QD. Figure
1c shows the sample structure, containing the doped back contacts,
the tunneling barrier t$_{b}$, the QD layer, the SRL (samples B and
C only), capping layer and metal Schottky contact. Finally, figure
1d shows the structure band diagram.

\begin{figure}
\includegraphics[%
  width=1.0\columnwidth]{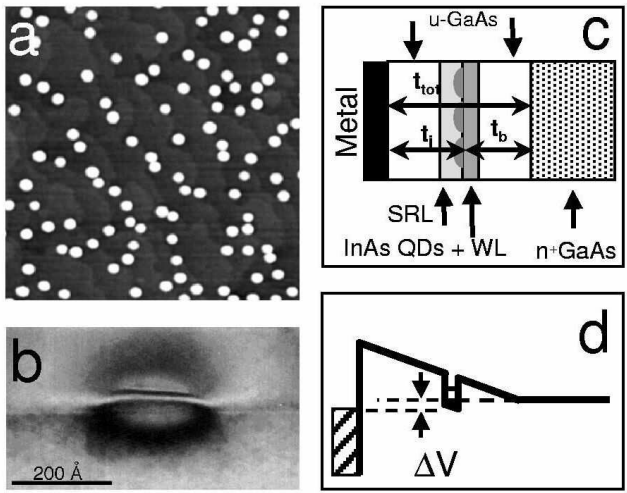}

\caption{\label{fig:1} Sample structure for capacitance spectroscopy and
g-factor assessment; (a) 1 $\mu m\times1\mu m$ AFM image on uncapped
InAs QDs; (b) XTEM image of a buried QD; (c) sample structure; and
(d) band diagram.}
\end{figure}

The Schottky diodes were processed over different portions of the
wafers, using standard lithography with AuGeNi ohmic back contacts
and 1 mm diameter Cr top Schottky gates.

\section{g-factor assessment}

\label{sec:2}

Insofar as spintronics applications are concerned, a key parameter
for spin manipulation is the g-factor. The g-factor of a QD determines
the radio frequency at which the spin of the electron trapped in the
dots would respond for a given static magnetic field. The electron
g-factor was evaluated using capacitance spectroscopy, which provides
direct information on the QD density of states (DOS).

The capacitance experiments were performed in a 15 T magnet at 2.2
K with lock-in amplifiers at frequencies ranging from 1 kHz to 10
kHz, and an ac amplitude of 4 mVrms superimposed on a varying dc bias.
The magnetic field was oriented parallel to the growth direction {[}001{]},
unless otherwise noticed. The signal/noise ratio for these experiments
was always above 10$^{5}$.

The contribution of the structure geometry in the measured capacitance
can be easily determined. By solving the Poisson equation for this
structure within the depletion approximation, and neglecting the contribution
of the QDs, we have:

\begin{equation}
C(V)=\frac{\epsilon_{0}\epsilon_{r}A}{2t_{tot}-\sqrt{t_{tot}^{2}-\frac{2\epsilon_{0}\epsilon_{r}}{qN_{D}}(\phi_{B}-V)}},\end{equation}

\noindent with t$_{tot}$ as the structure total thickness, $\epsilon_{0}$
and $\epsilon_{r}$ the dielectric constant and relative permissivity
of vacuum and medium, $A$ the sample area, $q$ the electronic charge,
$N_{D}$ the back contact doping concentration, $\phi_{B}$ the Schottky
barrier height, and $V$ the applied bias. This equation takes into
account depletion region effects in the back contact, which can be
neglected if the doping level is very high. All the capacitance traces
for high reverse biases were adjusted in order to obtain the sample
area and the doping concentration. The Schottky barrier height was
obtained from the flat band voltage as well as from current-voltage
characteristics obtained for similar devices. The doping levels and
sample area agreed to within 5\% of the nominal and measured values.
In the approximation of metallic contacts, the relation for the variation
of the chemical potential inside the QDs is $\Delta\mu=t_{b}\Delta V/t_{tot}$,
with the ratio $t_{b}/t_{tot}$ as the lever arm ratio. If one includes
the depletion region effects on the back contact, $\Delta\mu$ is
given by:

\begin{equation}
\Delta\mu=\frac{t_{b}+t_{tot}-\sqrt{t_{tot}^{2}-\frac{2\epsilon_{0}\epsilon_{r}}{qN_{D}}(\phi_{B}-V)}}{2t_{tot}-\sqrt{t_{tot}^{2}-\frac{2\epsilon_{0}\epsilon_{r}}{qN_{D}}(\phi_{B}-V)}}\Delta V.\end{equation}

It is important to note that this apparently small correction will
produce significant errors if the values for the doping levels as
determined by the fits of the capacitance data to equation (1) are
not used.

It has been demonstrated that InAs QDs can be modeled quite accurately
utilizing a lateral parabolic confinement approximation \cite{16},
for which the eigen-energies are known as a function of an external
magnetic field \cite{17}. Including the effect of the applied bias,
the Coulomb charge interaction between carriers sequentially added
to the QD ensemble, and the Zeeman splitting, the energies of states
with one and two electrons in the s shell are \cite{2}:

\begin{equation}
E_{s1}(B)=E_{z}+\hbar\sqrt{\omega_{0}^{2}+\frac{\omega_{c}^{2}}{4}}-\frac{|g|\beta B}{2}-\mu_{1}\end{equation}

and

\begin{equation}
E_{s2}(B)=2E_{z}+2\hbar\sqrt{\omega_{0}^{2}+\frac{\omega_{c}^{2}}{4}}+E_{CB}-2\mu_{2},\end{equation}

\noindent where $E_{z}$ is the vertical confining potential, $\hbar\omega_{0}$
the lateral confining potential characteristic energy, $\omega_{c}$
the cyclotron frequency, $|g|$ the g-factor modulus, $\beta$ the
Bohr's magneton, $B$ the magnetic field intensity, $E_{CB}$ the
Coulomb charging energy for the two electrons in the QD ground state,
and $\mu_{1}$ ($\mu_{2}$) the QD chemical potential at which the
first (second) electron tunnel into the ground state. One can relate
$\mu_{1}$ and $\mu_{2}$ with the applied bias through equation (2).

For the other shells one can easily expand the above expressions utilizing
the Fock-Darwin eigen-energies. Figure 2 shows capacitance spectra
with the background removed after fitting the CV characteristics at
large reverse bias with equation (1) for sample A. Figure 2a shows
the capacitance spectrum at zero magnetic field. Figure 2b shows a
grayscale map of the DOS dependence on the magnetic field, where white
means maximal density of states (about $1\times10^{11}$ cm$^{-2}$eV$^{-1}$)
and black zero. The calculated electronic spectra dependence on the
applied magnetic field given by equation (3) and its extensions to
higher angular momentum states (p$^{+}$, p$^{-}$, d) are shown.
WL denotes the wetting layer level with its corresponding diamagnetic
shift.

\begin{figure}
\includegraphics[%
  width=1.0\columnwidth]{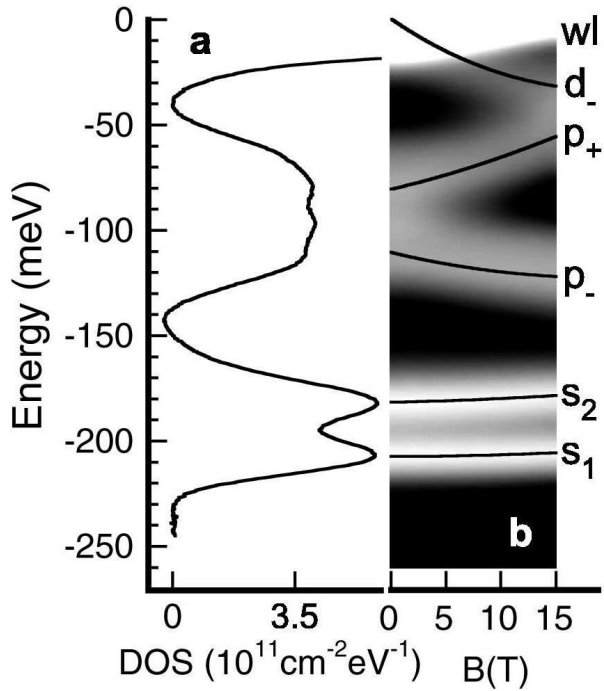}

\caption{\label{fig:2}Capacitance spectra for sample A. (a) Spectrum obtained
at $B=0$; the s and p shells can be completely loaded before the
wetting layer level. (b) QD density of states (DOS) dependence on
the magnetic field, with the grayscale representing 0 (black) and
$10^{11}$ cm$^{-2}$eV$^{-1}$ (white) for each of the confined levels.
In black, superimposed on the experimental data, the complete dependence
for the s1,2, p$^{-}$,$^{+}$ and d$^{-}$ states assuming a lateral
parabolic confining potential (Fock-Darwin states).}
\end{figure}

The voltage at which the first electron will tunnel into the QDs can
be found by $E_{0}$ = $E_{s1}$, $E_{0}$ ($E_{s1}$) being the total
energy of the system with no (one) electrons. For simplicity, we can
make $E_{0}=0$. The voltage for the second electron tunneling event
can be found in a similar fashion by equating $E_{s1}$ to $E_{s2}$.
From equations (3) and (4) and the above conditions, by measuring
the voltage difference from one to two electrons and utilizing equation
(2), we have:

\begin{equation}
\Delta\mu=\mu_{2}-\mu_{1}=E_{CB}+|g|\beta B.\end{equation}

From the dependence of $\Delta\mu$ on the magnetic field, one can
therefore obtain the g-factor modulus. Figure 3a shows the DOS for
s1 and s2 for 0, 7.5 T and 15 T for sample A. The peak spacing is
equal to $E_{CB}$ for $B$ = 0. Figure 3b the spin splitting dependence
on the magnetic field, obtained from $\Delta\mu-E_{CB}$ for several
fields and samples A and D. The g-factor modulus can be found from
the slopes of those curves and were $|g|=1.53\pm0.03$ and $1.50\pm0.04$
for samples A and D.

\begin{figure}
\includegraphics[%
  width=1.0\columnwidth]{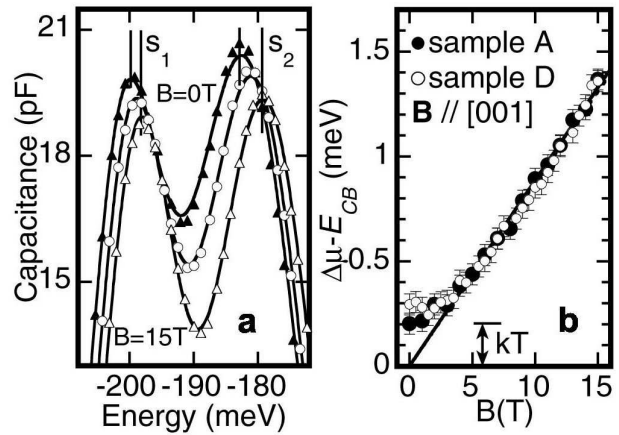}

\caption{\label{fig:3}(a) s1 and s2 peak dependence on the magnetic field
for $B=0$, 7.5 and 15 T. The shifts can be easily seen. (b) Dependence
on the magnetic field of the s1-s2 peak spacing excluding the Coulomb
charging energy contribution {[}equation (5){]}. The different symbols
refer to samples A and D, grown without the strain reducing layer
(SRL). The thermal energy $kT$ is also shown, which sets the magnetic
field scale for the spin splitting to be observed with capacitance
spectroscopy.}
\end{figure}

At low magnetic fields one can observe that the linear dependence
for the spin splitting disappears. This is a consequence of the thermal
energy being roughly the same as the spin-splitting energy for these
low magnetic fields (i.e, $|g|\beta B/kT<1$). For low fields, capacitance
spectroscopy probes quantum levels that can be occupied by either
spin orientation. For $|g|\beta B/kT>1$, the spin of the electrons
in the QDs becomes polarized and consequently a linear relationship
follows.

\section{g-factor tuning and engineering}

\label{sec:3}

The basic principle behind g-factor engineering and control lies on
the ability of either tailoring the barriers of a given quantum system
with different g-factor materials, and/or simultaneously swaying the
wave function in and out of the barriers. This comes about from the
fact that the effective g-factor of an electron or hole is

\begin{equation}
g_{eff}=\int_{A}g_{A}\Psi^{*}\Psi dV_{A}+\int_{B}g_{B}\Psi^{*}\Psi dV_{B},\end{equation}

\noindent with $g_{A}$ and $g_{B}$ the g-factors of the well material
(A) and barrier material (B). Therefore if one decreases the wave
function penetration inside a barrier the g-factor will change. This
was experimentally demonstrated by either applying an electric field
parallel to the growth direction \cite{11,12} or a magnetic field
perpendicular to the growth direction \cite{3}. For our structures,
the signature of a decrease of the tail of back contact and QD wave
functions ($\Psi_{BC}$, $\Psi_{s,p,d}$) is a decrease of the capacitive
signal of electrons trapped in the s1 state on the magnetic field
applied in the {[}110{]} direction. As expected, for stronger magnetic
fields the signal decreases (See Fig. 4a), indicating a compression
of the electron wave functions $\Psi_{s,p,d}$ and $\Psi_{BC}$ into
the QD and back contact. Figure 4b shows the capacitance peak spacing
excluding $E_{CB}$ ($\Delta\mu-E_{CB}$) as a function of the magnetic
field, exhibiting the non-linear dependence of the spin-splitting
and hence the g-factor change. The thermal energy is also represented,
indicating that for magnetic fields as high as 6 T, the spin splitting
is not clearly visible. Therefore, one can infer a g-factor smaller
than 0.5, which increases at stronger magnetic fields. This represents
the squeezing of the electron wave function into QDs, thereby altering
its in-plane g-factor component.

\begin{figure}
\includegraphics[%
  width=0.90\columnwidth]{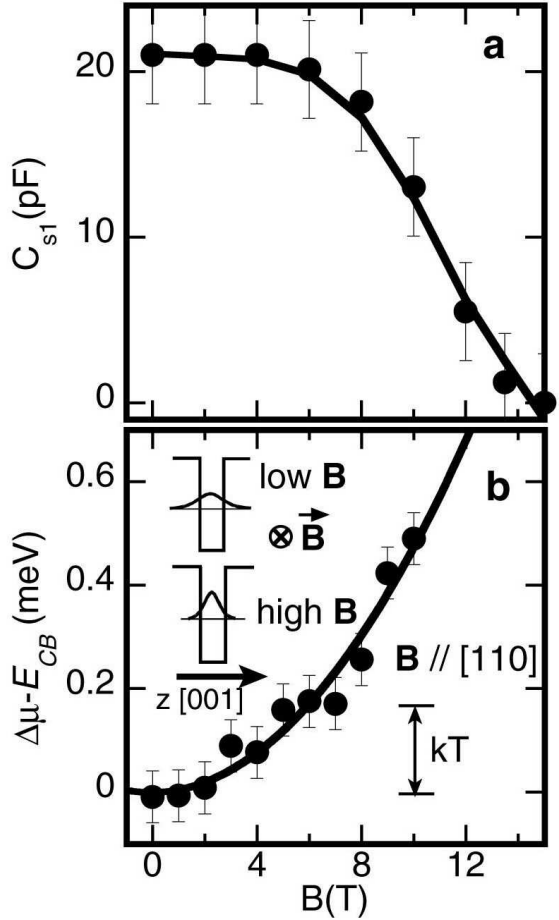}

\caption{\label{fig:4}(a) Capacitive signal of the electrons trapped in the
s1 state measured at a frequency of 5 kHz. Here, the intensity of
the signal is related to the hopping rate of the electrons between
s1 and the back contact, which is also proportional to the wavefunction
$\Psi_{s1}$ extent into the GaAs barrier. By applying a magnetic
field parallel to the barrier, one can squeeze the wavefunction inside
the QDs. (b) non-linear dependence of the spin-splitting, indicating
g-factor control via a magnetic field; the stronger the compression,
the larger the g-factor (inset).}
\end{figure}

\begin{table}
\begin{tabular}{ccccc}
\hline 
\noalign{\smallskip} Sample &
 $E_{0}$ (eV)/ $\Gamma$ (meV)&
 $\hbar\omega_{0}$ (meV)&
 $E_{CB}$ (meV)&
 g {[}001{]} \tabularnewline
\hline
 \noalign{\smallskip}A &
1.039/41.7 &
44 $\pm$1 &
25 $\pm$1 &
1.53 $\pm$0.03 \tabularnewline
B &
1.025/48.8 &
46 $\pm$1 &
20 $\pm$1 &
1.64 $\pm$0.02 \tabularnewline
C &
1.030/33.3 &
49 $\pm$1 &
20 $\pm$1 &
1.69 $\pm$0.02 \tabularnewline
D &
1.070/34.4 &
46 $\pm$2 &
18 $\pm$1 &
1.50 $\pm$0.04 \\
\noalign{\smallskip}\tabularnewline
\hline
\end{tabular}

\caption{\label{tab:1} Summary of the main parameters for the samples studied
in this work. Photoluminescence peak energy $E_{0}$ and corresponding
line width $\Gamma$, characteristic energy $\hbar\omega_{0}$, Coulomb
blockade energy $E_{CB}$ and g-factor along the {[}001{]} direction.~\protect \\
}
\end{table}

Samples B and C were grown with different capping layers, which according
to equation (6) should change the g-factor. Figure 5 and Table 1 summarizes
the main results of this paper. Figure 5a shows the grayscale maps
for the level dispersion for all samples. The Fock-Darwin levels can
be seen quite easily. From the s states, the characteristic energies
$\hbar\omega_{0}$ and Coulomb Blockade $E_{CB}$ can be extracted.
Those will influence the g-factor, as they are related to the extent
of the wave function. The Coulomb Energy can be evaluated in two ways:
a) by measuring the energy difference between states s1 and s2, and
b) from $\hbar\omega_{0}$, the characteristic length of the wave
function $l_{0}$ can be extracted as $l_{0}=\sqrt{\hbar^{2}/m^{*}\omega_{0}}$,
and from that $E_{CB}\simeq q^{2}/(4\pi\epsilon_{0}\epsilon_{r}l_{0}$).
These two different approaches agreed within 5\% for all samples but
sample D, supporting the parabolic confining potential assumption.
For sample D, a minor deviation of the parabolic confining potential
is observed, as the s-p-d level spacing is different from the values
obtained independently for each level dispersion with the magnetic
field (Fock-Darwin states). The presence of the d state at zero magnetic
field is a clear evidence of that. This reflects a larger size, and
can be verified as well by noting that the Coulomb charging energies
for sample A is larger than for sample D. Nevertheless, the value
for $\hbar\omega_{0}$ is about the same, meaning that although the
islands are different, the trapping potential for the s state is the
same.

\begin{figure}
\includegraphics[%
  width=1.0\columnwidth]{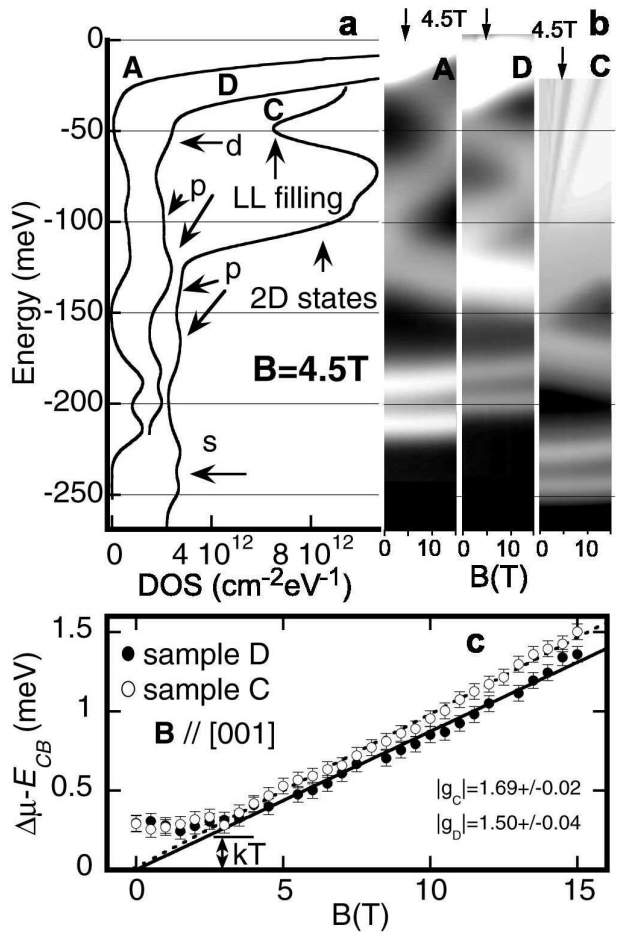}

\caption{\label{fig:5}(a) Capacitance spectra taken on samples A, C and D,
at 4.5 T, with the indicated 0D as well as 2D levels (Landau level
- LL); (b) corresponding grayscale representation of the DOS dependence
on the magnetic field, and (c) spin-splitting of the s state reflecting
g-factors differing in about 13\% for samples C and D.}
\end{figure}

From equation (5), the g-factor for all samples was determined, as
shown on Table 1. Samples A and D exhibited the same g-factor, within
the experimental uncertainties. This is an extremely important result,
evidencing the reproducibility of the procedure for the growth runs,
sample processing and measurement. In addition, differences in island
shape were not sufficient to change the g-factor. In contrast, the
effect of the SRL on the electronic properties and on the g-factor
of samples B and C is quite significant. The QD size does not change
appreciably, upon contrasting the energies $\hbar\omega_{0}$ and
Coulomb charging energies for all samples. The association of the
wetting layer with the SRL producing a thicker quantum well will lower
the energy for loading the corresponding 2D levels. A structure revealing
the loading of the Landau levels associated with the quantum well
can be seen by the dark straight lines on the capacitance spectra.
This quantum well prevents the observation of the d levels for both
samples, as they were always resonant with the 2D states and therefore
not accessible in this experiment. From the data presented here, one
can infer that the g-factor is connected to the characteristics of
the quantum well, in accord with equation 6. The g factor obtained
for samples B and C is systematically larger than for samples A and
D (up to 13\%). The differences for the g-factor for samples B and
C is about 5\%, which is experimentally accessible, however the precise
details for the differences in the quantum wells for the two samples
remain to be further investigated.

\section{Conclusions}

\label{sec:4}

The g-factor control and engineering proposed in this work have an
important impact in quantum information processing devices. It was
demonstrated that capacitance spectroscopy could be used to measure
g-factor of electrons trapped in the ground state of quantum dots.
The s-state level of the QDs revealed a robust character insofar as
the g-factor was concerned - QDs engineered with small shape differences
produced the same g-factor for the electrons on the s shell. This
important result addresses the issue of inhomogeneous broadening of
the size distribution of quantum dots and g-factor fluctuations -
it might not be as severe as one would anticipate. This contrasted
with structures with a capping strain reducing layer (SRL), intended
to modify the g-factor. By utilizing SRL structures as thin as 3 nm,
up to a 13\% difference was engineered on the g-factor. A wider range
of g-factors can be engineered for QDs, by varying the alloy composition
and thickness of the SRL and/or adding materials with different g-factor
sign (AlAs, for instance). The control of the g-factor by an external
applied magnetic field is yet another handle to be explored. In summary,
the g-factor engineering and control for single electrons were demonstrated
and represent a key step for solid state quantum information processing
devices based on QDs.

\noindent \textit{Acknowledgments.} This work was funded by CNPq,
FAPESP and HP Brazil. G.M.R wishes to acknowledge the TEM image by
D. Leonard, the use of the high-field magnet at the GPO group at the
IFGW, UNICAMP, the IFSC/USP for the access to the MBE apparatus, and
the technical support with the MBE by H. Arakaki, C. A. de Souza and
E. Marega.


\begin{thebibliography}{10}
\bibitem{1}D. P. DiVicenzo: e-print quant-ph 0002077. 
\bibitem{2}G. Medeiros-Ribeiro, F. G. Pikus, P. M. Petroff, A. L. Efros: Phys.
Rev. B \textbf{55}, 1568 (1997). 
\bibitem{3}G. Medeiros-Ribeiro, M. V. B. Pinheiro, V. L. Pimentel, E. Marega:
Appl. Phys. Lett. \textbf{80}, 4229 (2002). 
\bibitem{4}A. Abragam: \textit{Electron Paramagnetic Resonance of Transition
Ions} (Clarendon Press, Oxford 1970). 
\bibitem{5}M. Bayer, A. Kuther, A. Forchel, A. Gorburov, V. B. Timofeev, F. Sch\"{a}fer,
J. P. Reithmaier, T. L. Reinecke, S. N. Walck: Phys. Rev. Lett. \textbf{82},
1748 (1999). 
\bibitem{6}A. R. Go\~{n}i, H. Born, R. Heitz, A. Hoffman, C. Thomsen. F. Heinrichsdorff,
D. Bimberg: Jpn. J. Appl. Phys. \textbf{39}, 3907 (2000). 
\bibitem{7}J. G. Tischler, A. S. Bracker, D. Gammon, D. Park: Phys. Rev. B \textbf{66},
081310 (2002). 
\bibitem{8}S. Lindemann, T. Ihn, T. Heinzel, W. Zwerger, K. Ensslin, K. Maranowski,
A. C. Gossard: Phys. Rev. B \textbf{66}, 195314 (2002). 
\bibitem{9}J. R. Petta, D. C. Ralph: Phys. Rev. Lett. \textbf{89}, 156802 (2002). 
\bibitem{10}C. Durkan, M. E. Welland: Appl. Phys. Lett. \textbf{80}, 458 (2002). 
\bibitem{11}G. Salis, Y. Kato, K. Ensslin, D. C. Driscoll, A. C. Gossard, J. Levy,
D. D. Awschalom: Nature \textbf{414}, 619 (2001). 
\bibitem{12}Y. Kato, R. C. Myers, D. C. Driscoll, A. C. Gossard, J. Levy, D. D.
Awschalom: Science \textbf{299}, 1201 (2003). 
\bibitem{13}H. Kosaka, A. A. Kiselev, F. A. Baron, K. W. Kim, E. Yablonovitch:
Electronics Letters \textbf{37}, 464 (2001). 
\bibitem{14}A. A. Kiselev, K. W. Kim, E. Yablonovitch: Appl. Phys. Lett. \textbf{80},
2857 (2002). 
\bibitem{15}K. Nishi, H. Saito, S. Sugou, J.-S. Lee: Appl. Phys. Lett. \textbf{74},
1111 (1999). 
\bibitem{16}M. Fricke, A. Lorke, J. P. Kotthaus, G. Medeiros-Ribeiro, P. M. Petroff:
Europhys. Lett. \textbf{36}, 197 (1996); R. J. Warburton, C. S. Durr,
K. Karrai, J. P. Kotthaus, G. Medeiros-Ribeiro, P. M. Petroff: Phys.
Rev. Lett. \textbf{79}, 5282 (1997); P. Hawrilak, G. A. Narvaez, M.
Bayer, A. Forchel, Phys. Rev. Lett. \textbf{85}, 389 (2000). 
\bibitem{17}V. Fock: Z. Phys. \textbf{47}, 446 (1928). \end{thebibliography}
\end{document}